\documentclass[fleqn,usenatbib]{mnras}

\usepackage{newtxtext,newtxmath}

\usepackage[T1]{fontenc}
\usepackage{adjustbox}
\usepackage{tablefootnote}

\DeclareRobustCommand{\VAN}[3]{#2}
\let\VANthebibliography\thebibliography
\def\thebibliography{\DeclareRobustCommand{\VAN}[3]{##3}\VANthebibliography}

\usepackage{graphicx}	
\usepackage{adjustbox}
\usepackage{arydshln}
\usepackage{longtable}
\usepackage{lscape}

\defcitealias{Bhattacharya24}{Bh+24}
\defcitealias{MD14}{MD14}

\title[AUDFs III: UVLF evolution from z $\sim$ 0.8--0.4]{The AstroSat UV Deep Field South. III. Evolution of the UV Luminosity Function and Luminosity Density from z $\sim$ 0.8 -- 0.4}

\author[Souradeep Bhattacharya and Kanak Saha]{
Souradeep Bhattacharya,$^{1, 2}$\thanks{E-mail: s.bhattacharya3@herts.ac.uk} and
Kanak Saha$^{2}$
\\
$^{1}$Centre for Astrophysics Research, Department of Physics, Astronomy and Mathematics, University of Hertfordshire, Hatfield AL10 9AB, UK\\
$^{2}$Inter University Centre for Astronomy and Astrophysics, Ganeshkhind, Post Bag 4, Pune 411007, India
}

\date{Accepted 2025 April 14;  Received 2025 April 01; in original form 2024 October 04}

\pubyear{2025}

\begin{document}
\label{firstpage}
\pagerange{\pageref{firstpage}--\pageref{lastpage}}
\maketitle

\begin{abstract}
We characterise the rest-frame 1500~\AA~UV luminosity Function (UVLF) from deep AstroSat/UVIT F154W and N242W imaging in the Great Observatories Origins Survey South (GOODS-S) deep field. The UVLFs are constructed and subsequently characterised with fitted Schechter function parameters from FUV observations at z$<0.13$ and NUV observations in seven redshift bins in z$\sim$0.8--0.4. The UVLF slope ($\alpha$) and characteristic magnitude ($M^*$) are consistent with previous determinations for this redshift range based on AstroSat/UVIT GOODS-North observations, as well as with those from \textit{Galaxy evolution Explorer} and \textit{Hubble Space Telescope} observations. However, differences in the normalisation factor ($\phi_{*}$) are present for UVLFs for some redshift bins. We compute the UV luminosity density, $\rho_{\rm UV}$, combining our determined UVLF parameters with literature determinations out to z$\sim10$. The $\rho_{\rm UV}$ trend with redshift implies the rapid increase in cosmic star formation till its peak at z$\sim3$ (cosmic noon) followed by a slow decline till present day. Both the initial increase in cosmic star formation and subsequent decline are found to be more rapid than previous determinations.
\end{abstract}





\section{Introduction}
\label{sec:intro}

Quantifying the distribution of luminosities within a given population, galaxy luminosity functions are key probes into the underlying physical processes that govern galaxy formation and evolution (see review by \citealt{Johnston11}). The rest-frame 1500~\AA~UV luminosity function (UVLF), sensitive to ongoing star formation in galaxies \citep{wyder05}, offers insights into the star formation activity of galaxies across cosmic time.

The UVLF is typically characterised within a given redshift range with the classic Schechter function \citep{Schechter76} parameterised in the following manner as a function of magnitude M:
\begin{equation}
    \phi(M) dM = \frac{ln(10)}{2.5}~\phi_{*}~(10^{0.4\Delta M})^{\alpha + 1}e^{-10^{0.4\Delta M}}dM
\end{equation}
where $\Delta M = M^* - M$, $\phi_{*}$ is  a normalization factor which defines the overall density of galaxies (number per cubic Mpc),  $M^*$ is the characteristic galaxy magnitude, and $\alpha$ defines the faint-end slope (typically negative implying large numbers of relatively passive galaxies with low luminosities).

The UVLF has been extensively characterized at z$\sim$1--9 from \textit{Hubble Space Telescope} (HST) surveys in deep-fields\footnote{Sensitive observations over a small region of the sky that provide unique opportunities to detect and analyze faint, low-luminosity galaxies.} \citep[e.g.][]{oesch10, Alavi16, Bouwens22, Finkelstein15} and at z$>$8 from recent \textit{James Webb Space Telescope} (JWST) deep-field observations \citep[e.g.][]{Bouwens23,Finkelstein24}. However, at z$<1$ which spans a wide range of ages ($\sim$8~Gyr), direct UVLF characterisation has been more limited to relatively bright sources from \textit{Galaxy evolution Explorer} (GALEX; \citealt{Arnouts05,wyder05}), \textit{Swift UV/Optical Telescope} (UVOT; \citealt{Hagen15}) and \textit{XMM-Newton} Optical Monitor (XMM-OM; \citealt{Page21,Sharma22a,Sharma24}) observations, as well as earlier ones from balloon-borne observations \citep{Treyer98,Sullivan00}. The preferential observations of bright sources is due to relatively low angular resolution of near-UV (NUV) channels for the aforementioned telescopes (FWHM$\rm_{NUV, GALEX}=5.3^{\prime\prime}$; FWHM$\rm_{NUV, UVOT}=2.5^{\prime\prime}$; FWHM$\rm_{NUV, XMM-OM}=2^{\prime\prime}$) adversely affecting the detection of faint sources in crowded deep fields. 

The UV Imaging Telescope (UVIT; \citealt{Kumar12}) on board \textit{AstroSat} \citep{singh14} recently carried out higher angular-resolution (FWHM$\rm_{F154W}=1.18^{\prime\prime}$, FWHM$\rm_{N242W}=1.11^{\prime\prime}$) UV observations of the Great Observatories Origins Deep Survey - North and South deep fields (GOODS-N and GOODS-S respectively; \citealt{Giavalisco04}). The observations of the GOODS-N field are described in \citet{Mondal23}. In \citet[][hereafter Bh+24]{Bhattacharya24}, we presented UVLFs for z$<=0.13$ for a single bin and for $0.378<$~z~$<0.768$ for 7 bins (bin size $\sim0.055$ in z), thereby characterising the UVLF in the GOODS-N field with unprecedented resolution in redshift and thus probing the variation of the fitted UVLF parameters over  $\sim$2.7~Gyr in age. We found that $\alpha$ was at its steepest at z$\sim$0.63, potentially implying highest star-formation at this instant with galaxies being relatively more passive before and after this time within the probed redshift range.

Our recent UV observations of the GOODS-S deep field with AstroSat/UVIT \citep{Saha24}, deeper than those in the GOODS-N field allow us to probe the behavior of the fitted UVLF parameters in an additional field, and draw potentially more universally applicable inferences on cosmic star-formation at z$\sim$0.4--0.8. Furthermore, from the UVLF parameters thus determined at low-z and given the availability of fitted UVLF parameters out to z$\sim$10, the behaviour of cosmic star-formation may be analysed from computing the UV luminosity density \citep[e.g.][]{oesch10,Alavi16} over this large redshift range.

We describe the AstroSat/UVIT GOODS-S data and associated photometric redshift utilised in this work in Section~\ref{sec:data}. The constructed UVLFs are described in Section~\ref{sec:uvlf}. We discuss the consistency of the fitted UVLF parameters with past determinations in Section~\ref{sec:disc} along with the evolution of the UV luminosity density. We conclude in Section~\ref{sec:conc}.

Throughout this paper, we use the standard cosmology ($\rm\Omega_{m}$ = 0.3, $\rm\Omega_{\Lambda}$ = 0.7 with $\rm H_{0}$ = 70 km~s$^{-1}$~Mpc$^{-1}$). 
Magnitudes are given in the AB system \citep{Oke74}.

\begin{table*}
\caption{Fitted UVLF Parameters. Column 1: Central redshift of each bin; Column 2: Range of redshifts spanned by each bin; Column 3: No. of galaxies in each bin considered for UVLF fitting; Column 4: No. of AGN removed in each bin; Columns 5--7: Fitted Schechter function parameters; Column 8: Logarithm of luminosity density.}
\centering
\adjustbox{max width=\textwidth}{
\begin{tabular}{cccccccc}
\hline
$\rm z_{mean}$ & $\rm z_{range}$ & No. of & No. of & $\phi_*$ & M* &
$\alpha$ &
log($\rho_{\rm UV}$)\\
 &  & galaxies & AGN & $10^{-3}$ mag$^{-1}$ Mpc$^{-3}$ & mag &  & erg/s/Hz/Mpc$^3$\\
\hline
\multicolumn{8}{c}{AUDFs FUV}\\
\hline
0.1 & 0.01 -- 0.13 & 189 & 11 & $ 0.91 \pm 0.3 $ & $ -18.03 $ & $ -1.65 \pm 0.08 $ & $25.17 \pm 0.14$\\
\hline
\multicolumn{8}{c}{AUDFs NUV}\\
\hline
0.41 & 0.378 -- 0.434 & 189 & 5 & $ 1.85 \pm 1.54 $ & $ -18.51 \pm 0.68 $ & $ -1.54 \pm 0.2 $& $25.53 \pm 0.45$\\
0.46 & 0.434 -- 0.489 & 157 & 3 & $ 4.19 \pm 2.35 $ & $ -17.97 \pm 0.55 $ & $ -1.13 \pm 0.23 $& $25.43 \pm 0.33$\\
0.52 & 0.489 -- 0.545 & 274 & 14 & $ 5.11 \pm 4.29 $ & $ -18.36 \pm 0.6 $ & $ -1.27 \pm 0.25 $& $25.73 \pm 0.44$\\
0.57 & 0.545 -- 0.601 & 269 & 13 & $ 7.55 \pm 4.53 $ & $ -17.94 \pm 0.42 $ & $ -1.32 \pm 0.28 $& $25.76 \pm 0.31$\\
0.63 & 0.601 -- 0.657 & 320 & 16 & $ 4.23 \pm 3.12 $ & $ -18.79 \pm 0.56 $ & $ -1.37 \pm 0.27 $& $25.87 \pm 0.39$\\
0.68 & 0.657 -- 0.712 & 347 & 36 & $ 9.73 \pm 4.1 $ & $ -18.59 \pm 0.39 $ & $ -0.96 \pm 0.27 $& $25.99 \pm 0.46$\\
0.74 & 0.712 -- 0.768 & 387 & 33 & $ 10.94 \pm 1.57 $ & $ -18.55 \pm 0.13 $ & $ -0.94 \pm 0.1 $& $26.03 \pm 0.08$\\
\hline
\multicolumn{8}{c}{HDUV-S}\\
\hline
0.71 & 0.679 -- 0.734 & 126 & 8 & $ 1.91 \pm 0.99 $ & $ -19.03 \pm 0.46 $ & $ -1.33 \pm 0.16 $ & $25.57 \pm 0.35$\\
0.76 & 0.734 -- 0.79 & 153 & 11 & $ 2.39 \pm 1.0 $ & $ -18.98 \pm 0.4 $ & $ -1.15 \pm 0.16 $ & $25.6 \pm 0.24$\\
\hline
\end{tabular}
\label{tab:uvlf}
}
\end{table*}


\section{Galaxy Catalogue Data} 
\label{sec:data}

In \citet{Saha24}, we present deep UV imaging observations of the GOODS-S field with AstroSat/UVIT (AstroSat UV Deep Field South, hereafter AUDFs), using one FUV (F154W) and one NUV filter (N242W). The deep and shallow parts of AUDFs have exposure times of $\sim 62000$~s ($\sim 64000$~s) and $\sim31000$~s ($\sim34000$~s) respectively in F154W (N242W). This imaging covered $\sim$236 sq. arcmin, including $\sim$193 sq. arcmin covered by the HST CANDELS survey with multiple HST filters \citep{Grogin11} and 43.4 sq. arcmin spanned by the Hubble Deep UV Legacy Survey in the GOODS-S field (HDUV-S) with HST F275W and F336W filters \citep{Oesch18}. 

As was the case for AUDFn (see \citetalias{Bhattacharya24} for details), we utilize only those F154W and N242W sources catalogued by \citet{Saha24} that have a single unique HST counterpart within $1.4''$. We additionally restrict ourselves to only those sources that are brighter than the 50\% completeness limits of 27.63 (27.3) and 27.04 (26.73)~mag for the deep (shallow) N242W and F154W images respectively. The redshifts for our galaxies were adopted from the z$\rm_{best}$ value of the CANDELS Photometric Redshift Catalog \citep{Kodra23}.  We utilise only those galaxies to construct the rest-frame 1500~\AA~UVLFs that are at z$<=0.13$ and $0.378<$~z~$<0.768$ in the F154W and N242W filters respectively, where they are most sensitive to the redshifted 1500~\AA~ as per their filter transmission functions \citep{Tandon20}. We thus select 200 and 2067 sources with only a single HST counterpart in the F154W and N242W images respectively within these redshift ranges. 

Additionally, we utilise the HST F275W photometric catalogue from HDUV-S \citep{Oesch18} to construct UVLFs for galaxies at $0.679<$~z~$<0.79$ as a consistency check to the N242W based UVLFs. There are 298 galaxies in this redshift range observed in F275W filter down to the 5$\sigma$ detection limit of 27.6~mag \citep{Oesch18}.
 
The presence of any active galactic nuclei (AGN), especially bright ones (M$\rm_{1500}<-19$~mag), may affect the determined parameters of the UVLF. The presence of faint AGN is expected to only negligibly affect the fitted UVLF parameters at this redshift range \citepalias{Bhattacharya24}. \citet{Lyu22} presented a catalogue of 901 AGN identified in the GOODS-S region using diagnostics covering X-ray to radio observations. Of these 11, 120 and 19 have counterparts in the redshift ranges of interest for the aforementioned F154W, N242W and F275W sources respectively. As we found many of these (none in F154W but 31 in N242W, and 5 in F275W) to have M$\rm_{1500}<-19$~mag (unlike in AUDFn; \citetalias{Bhattacharya24}), we have removed these known AGN from our sample before fitting the UVLFs.


\begin{figure}
\centering
\includegraphics[width=0.9\columnwidth]{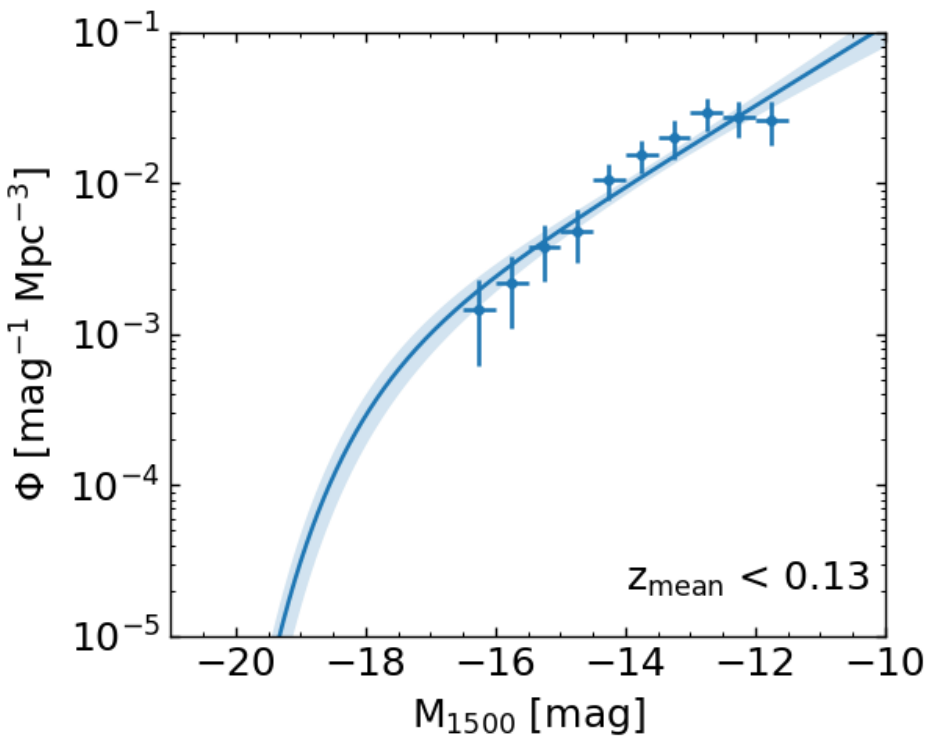}
\caption{The 1500~\AA~LF at z~$<0.13$ from the F154W imaging of the GOODS-S field. The Schechter function fit is marked with a solid blue line, with the 1~$\rm\sigma$ uncertainty shaded.}
\label{fig:fuv}
\end{figure}

\begin{figure*}
\includegraphics[width=0.9\textwidth]{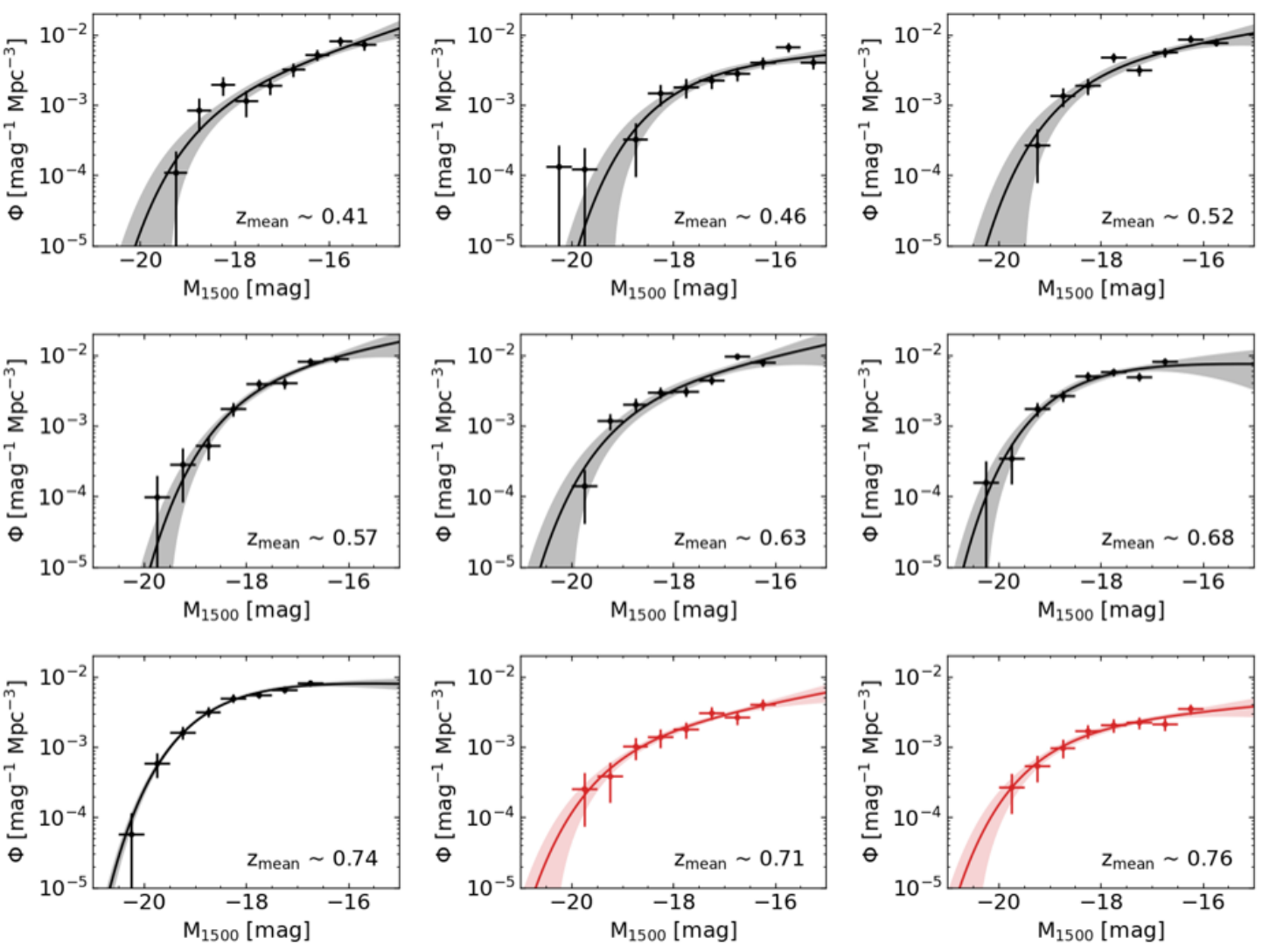}
\caption{The 1500~\AA~LF from AstroSat/N242W images of the GOODS-S field in seven different redshift bins are marked in black. The z$\rm_{mean}$ of each bin is also noted. The Schechter function fits for each bin is marked with a solid black line, with the 1~$\rm\sigma$ uncertainty shaded in grey. The center and right panels of the bottom row show the 1500~\AA~LF from HST F275W images of the GOODS-S field in two different redshift bins, marked in red. The Schechter function fit for each bin is marked with a solid red line, with the 1~$\rm\sigma$ uncertainty shaded. }
\label{fig:uvlf}
\end{figure*}

\begin{figure}
\centering
\includegraphics[width=0.9\columnwidth]{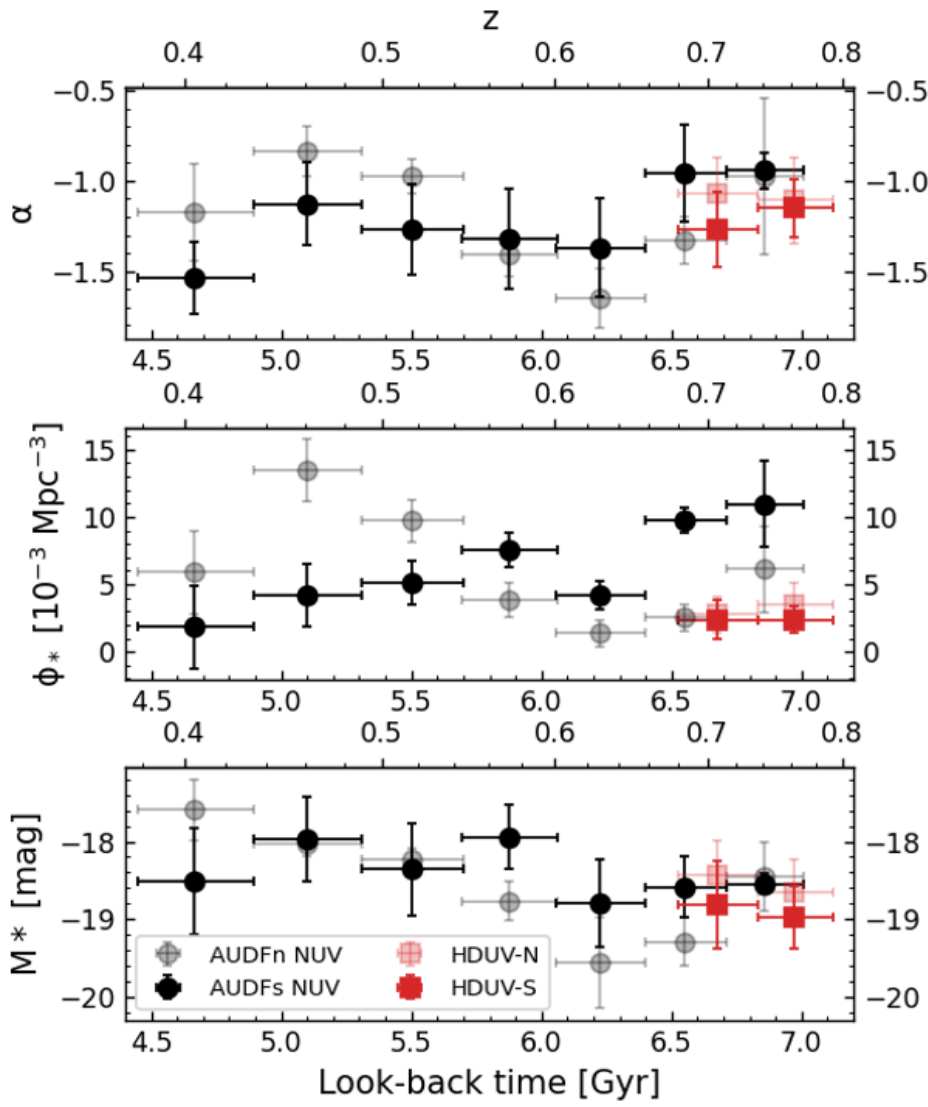}
\caption{Evolution of the fitted Schechter function parameters with look-back time compared for GOODS-S (this work) and GOODS-N \citepalias{Bhattacharya24}.}
\label{fig:compare}
\end{figure}

\section{The UV (1500~\AA) Luminosity Functions} 
\label{sec:uvlf}

The 1500~\AA~LF for z$<=0.13$ for a single bin, $0.378<$~z~$<0.768$ for 7 bins, and $0.679<$~z~$<0.79$ for two bins are fitted from the aforementioned observed number of galaxies in the F154W, N242W and F275W filters respectively. The procedure followed is the same as for AUDFn \citepalias{Bhattacharya24} which we describe briefly as follows:
\begin{enumerate}
    \item We carry out completeness correction based on the recovery fraction of sources as a function of apparent magnitude (see Figure 14 in \citealt{Saha24}) that was done using artificial source injection \citep[e.g.][]{Bhattacharya17,Bhattacharya19} for each image.
    \item Further corrections are applied to account for the fraction of total identified sources that have a unique HST counterpart as well as small K-corrections are derived from appropriate best-fit templates.
    \item We compute the luminosity distance of each galaxy, remove the known AGN, and construct the UVLF whose binned representation follows the method of \citet{Page00}.
    \item The UVLF is then fitted with the classic Schechter function \citep{Schechter76} using a maximum likelihood estimator fit.
\end{enumerate}  

For the F154W based UVLF at z$<=0.13$ (see Figure~\ref{fig:fuv}), galaxies are observed down to M$_{1500}$=-11.5~mag but there is a lack of bright galaxies (M$_{1500}>$-16.5~mag). We thus fix $M^*=-18.03$ (same value as \citealt{Arnouts05} from GALEX FUV data). The fitted $\phi_{*}$ and $\alpha$ values for the best-fit Schechter function are noted in Table~\ref{tab:uvlf}. 

To construct the N242W based UVLFs at $0.378<$~z~$<0.768$, the observed galaxies are divided into 7 equal bins in redshift ($\rm\Delta z = 0.055$). The mean redshift of each bin, z$\rm_{mean}$, the range of redshifts as well as the number of galaxies and AGN in each bin are noted in Table~\ref{tab:uvlf}. The binned representation of the UVLFs for each redshift bin is shown in Figure~\ref{fig:uvlf}. The Schechter function is fitted to the UVLF in each redshift bin keeping $\alpha$, $\phi_{*}$ and M$^{*}$ as free parameters. The best-fit Schechter function parameters are noted in Table~\ref{tab:uvlf}. 

The galaxies from the HDUV-S sample \citep{Oesch18} are divided into two redshift bins ($\rm\Delta z = 0.055$) at $0.679<$~z~$<0.79$. The binned representation of these UVLFs is shown in red in Figure~\ref{fig:uvlf}, but without any completeness correction (see Section 3.4 in \citetalias{Bhattacharya24} for details). Galaxies are observed down to M$_{1500}$=-16~mag for both the redshift bins, deeper than in the N242W data at the same redshift. The Schechter function is fitted to the UVLF for both redshift bins keeping $\alpha$, $\phi_{*}$ and M$^{*}$ as free parameters. The best-fit parameters are noted in Table~\ref{tab:uvlf}.


\begin{figure*}
\centering
\includegraphics[width=\textwidth]{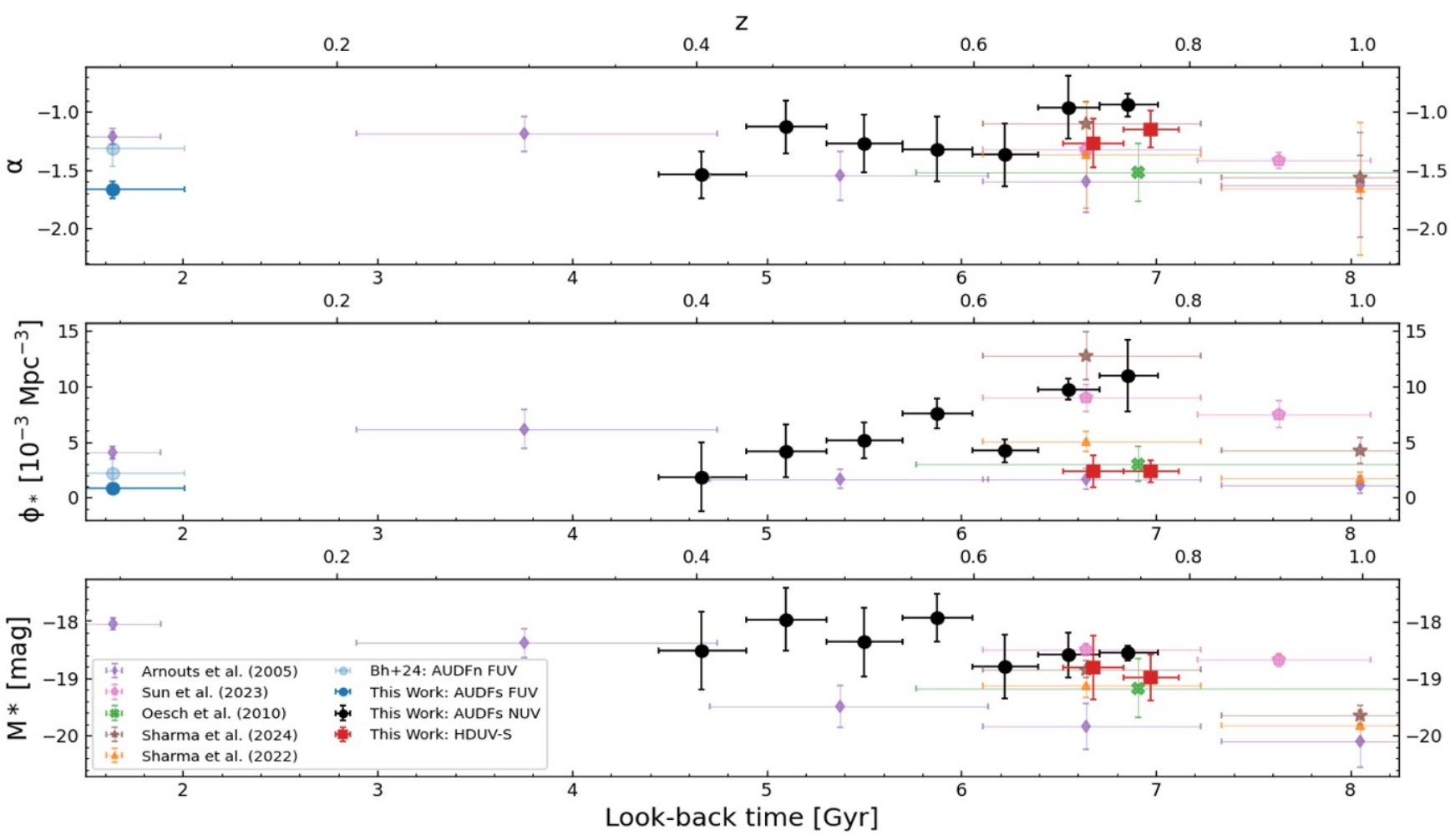}
\caption{Evolution of the fitted Schechter function parameters with look-back time (bottom axes) and redshift (top axes). The parameters from this work and from literature estimates are marked separately.} 
\label{fig:param}
\end{figure*}

\section{Discussion and Inference} 
\label{sec:disc}

\subsection{Consistency with AUDFn} 
\label{sec:compare}

Figure~\ref{fig:compare} shows the fitted Schechter function parameters determined in this work from Astrosat/UVIT N242W and HST F275W galaxies, compared with those determined in the GOODS-N region for galaxies identified with the same filters by \citetalias{Bhattacharya24}. For the HST F275W galaxies, UVLFs for HDUV-S and HDUV-N (in the GOODS North region) give consistent fitted Schechter function parameters within error for both redshift bins.For the N242W galaxies for $\alpha$ and M$^{*}$, there is consistency within error in these determined parameters across the seven redshift bins. 

In the redshift range $0.378<$~z~$<0.768$ probed here, the lowest fitted $\alpha$ value is at z$\rm_{mean}$=0.41. For GOODS-N, the galaxies in this redshift bin were found to have a lower mean stellar mass ($\sim10^{8.5}$~M$_{\odot}$) than those in the higher redshift bins with mean stellar mass $\sim10^{9}$~M$_{\odot}$ \citepalias{Bhattacharya24}. This is also expected to be true for GOODS-S and explains the lower fitted $\alpha$ value, given the presence of star-forming lower mass galaxies. 

The apparent dip in $\alpha$ at z$\rm_{mean}=0.63$ identified in the UVLF parameters fitted to GOODS-N galaxies \citepalias{Bhattacharya24} is also seen for GOODS-S (see Figure~\ref{fig:compare}). Comparing the distribution of the determined $\alpha$ values in this work for the 7 redshift bins of AUDFs NUV and 2 redshift bins of HDUV-S (see Table~\ref{tab:uvlf}) with a normal distribution (having the same mean, <$\alpha$>=-1.22, and standard deviation, $\sigma_{\alpha}$=0.18, as the 9 $\alpha$ values) using a Kolmogorov–Smirnov test \citep{kstest}, we obtain a p-value$>0.05$ (KS-statistic=0.155, p-value=0.96) implying the $\alpha$ values may be consistent with a normal distribution. We thus find that while the apparent dip in $\alpha$ at z$\rm_{mean}=0.63$ may be physical (as this is observed now for both GOODS-S and GOODS-N deep fields), the dip itself is not statistically significant based on the currently available data.

Variations with redshift, similar to that of $\alpha$, are not seen for $\phi_{*}$ and M$^{*}$, unlike the case for AUDFn (see Figure~\ref{fig:compare}), though a seemingly increasing trend is seen for $\phi_{*}$ with z, with a noticeable dip at z$\rm_{mean}=0.63$. The deeper survey likely helps us break the degeneracies between the different fitted Schechter parameters. There is however a discrepancy in the fitted $\phi_{*}$ values with those determined from fitting AUDFn galaxies being higher than those determined from fitting AUDFs galaxies at z$\rm_{mean}=0.46$~\&~$0.52$, while the opposite is true for z$\rm_{mean}=0.57$~--~$0.74$. 

As $\phi_{*}$ is the normalisation term for the Schechter function fit, it relates closely to the number of galaxies observed in any redshift window. If we restrict the GOODS-S N242W galaxies to only those brighter than 27.05~mag (50\% completeness limit of the shallower AUDFn NUV images; \citetalias{Bhattacharya24}), the numbers of galaxies are similar to that in GOODS-N field for the z$\rm_{mean}=0.57$~--~$0.74$ redshift bins. The additional survey depth for AUDFs reveals a higher number of fainter galaxies at these redshift bins than anticipated from their UVLF fits, leading to higher fitted $\phi_{*}$ values, but this is not as impactful for lower redshift bins. 

It is surprising to find that in the redshift bin with z$\rm_{mean}=0.46$, there are almost $\sim50$~\% more galaxies in GOODS-N than GOODS-S, while the former is the shallower surveyed field. \citet{Yantovski-Barth24} have identified a number of galaxy clusters in the DESI legacy survey \citep{Dey19} images. There is one at z$=0.4687$ in the GOODS-N field and one more at z$=0.4834$ centered just beyond this field. Members of both these clusters may be inflating the identified number of galaxies, and hence the determined $\phi_{*}$ in the z$\rm_{mean}=0.46$ redshift bin for GOODS-N. While most cluster members would be passive, it is surprising to see such a noticeable effect in the numbers of GOODS-N NUV sources, which are all expected to be star-forming. We will explore clusters in the AUDFs and AUDFn fields in detail in a future publication.   

\subsection{Consistency with past determinations of the UVLF} 
\label{sec:lit}

Figure~\ref{fig:param} shows the variation of the three Schechter function parameters, $\alpha$, $\phi_{*}$ and M$^{*}$, with redshift in this work, as well as other direct determinations of these parameters up to z$\sim1$ from the literature. At the lowest redshift (z$\sim0.1$) for the UVLF from AstroSat/F154W,  the determined $\alpha$ value is more negative than that determined in the GOODS-N field by \citetalias{Bhattacharya24} and by \citet{Arnouts05} from GALEX. However, the $\phi_{*}$ values are consistent across all three surveys. 

The $\alpha$ and M$^{*}$ values determined from HDUV-S are consistent with that determined for the highest redshift bin from AUDFs (see Table~\ref{tab:uvlf}) but the $\phi_{*}$ value is lower. This is likely the same effect of the shallower survey depth that also affected the $\phi_{*}$ values of the highest redshift bins in AUDFn. Combined with the lack of completeness correction, this effect seems more pronounced leading to low $\phi_{*}$ values. Using the same HST F275W and also the F435W filters but for a larger area of $\sim426$ sq. arcmin and a 5$\sigma$ detection limit of 27 mag, \citet{Sun23} presented the UVLF at z$\sim0.6$--$0.8$ and z$\sim0.8$--$1$ respectively. For all three fitted Schechter function parameters, our determined values at z$\rm_{mean}=0.68$~\&~$0.74$ are in good agreement with their determination at z$\sim0.6$--$0.8$ (see Figure~\ref{fig:param}). 

Additionally, the XMM-OM based UVLF parameters for z= 0.6--0.8 determined for the COSMOS field by \citet{Sharma24} are also consistent with all three fitted Schechter function parameters at the same highest redshift bins (see Figure~\ref{fig:param}). Other parameter determinations from fitting the UVLF to relatively shallower data from different works \citep{Arnouts05,oesch10,Sharma22a}, still results in some parameters that agree within error with that from AUDFs, though the $\phi_{*}$ values are generally lower for these works (see Figure~\ref{fig:param}). The fitted M$^{*}$ values from \citet{Arnouts05} are also much lower.

\begin{figure*}
\centering
\includegraphics[width=\textwidth]{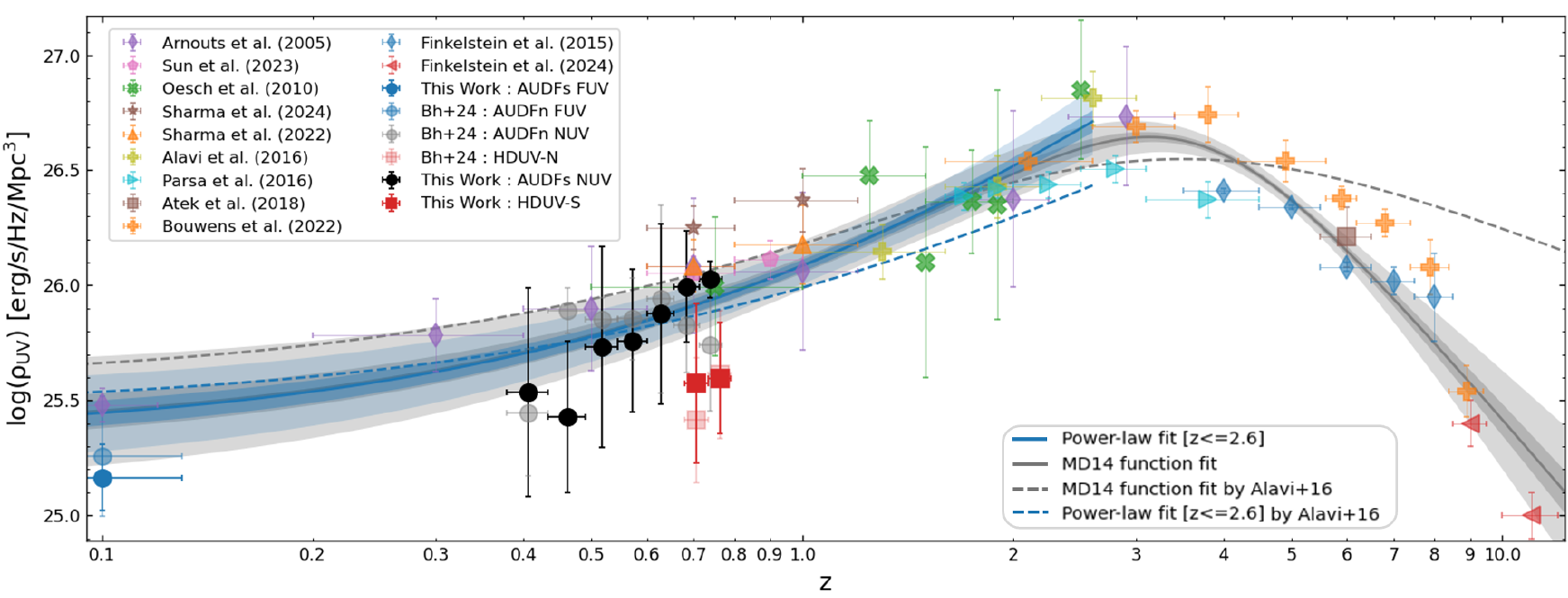}
\caption{Evolution of the fitted UV luminosity density with redshift (shown in log scale). The best-fit power-law out to z$=2.6$ is marked in blue with its 1~$\sigma$ and 3~$\sigma$ bounds marked with darker and lighter shaded regions. The best-fit \citetalias{MD14} function for the entire redshift range is similarly marked in grey. The same functions with fitted parameters previously determined by \citet{Alavi16} are also marked.}
\label{fig:lumden}
\end{figure*}

\subsection{Evolution of the UV luminosity density with redshift} 
\label{sec:evo}

We determine the UV luminosity density for each UVLF determined in this work for the GOODS-S region (by integrating them down to M$_{1500}=-10$). Its logarithm values, log($\rho_{\rm UV}$), are noted in Table~\ref{tab:uvlf} and plotted against redshift in Figure~\ref{fig:lumden}. We note that the trend in $\alpha$ (Figure~\ref{fig:param}), dipping at z$\rm_{mean}$=0.63 for both GOODS-S (this work) and GOODS-N fields \citepalias{Bhattacharya24}, does not reflect on the log($\rho_{\rm UV}$) distribution with redshift, which shows a increasing trend with increasing redshift. The redshift variations of M$^{*}$ and $\phi_{*}$ thus balance the trend in $\alpha$ such that log($\rho_{\rm UV}$) trend with redshift does not show any increased star-formation at z$\rm_{mean}$=0.63\footnote{log($\rho_{\rm UV}$) is still the highest at z$\rm_{mean}$=0.63 for the GOODS-N field (see Figure~\ref{fig:lumden}) as noted in \citetalias{Bhattacharya24} but that is a reflection of the lower $\phi_{*}$ values at higher z for the shallower GOODS-N survey, as discussed in Section~\ref{sec:compare}.}. The potential of such a short-lived instance of increased star-formation was discussed in detail in \citetalias{Bhattacharya24}. 

We additionally determine log($\rho_{\rm UV}$) for the lower redshift UVLFs presented by other aforementioned works from their reported fitted parameters, including those from \citet{oesch10} and \citet{Arnouts05} that go out to z$\sim3$. We also determine log($\rho_{\rm UV}$) from the reported parameters of some other works where UVLFs are determined at higher redshifts, i.e., \citet{Alavi16}: z$\sim$1--3, \citet{Parsa16}: z$\sim$1.8--4, \citet{Atek18}: z$\sim$6, \citet{Finkelstein15}: z$\sim$4--8, \citet{Finkelstein24}: z$\sim$9--11 and \citet{Bouwens22}: z$\sim$3--8. While a number of authors have determined the 1500~\AA~LF at z$>3$ in recent years, the aforementioned works are those that determined the UVLF down to the greatest observed depths for their presented redshifts. We note that the UVLF beyond z$\sim8$ is not well-constrained \citep{Bouwens23}. 

Figure~\ref{fig:lumden} shows the redshift evolution of log($\rho_{\rm UV}$). The highest log($\rho_{\rm UV}$) values are seen at z$\sim$2.5 \citep{Arnouts05,oesch10,Alavi16} with decreasing log($\rho_{\rm UV}$) values computed at both higher and lower redshift values as we move away from cosmic noon. Akin to previous authors \citep{oesch10,Alavi16}, we fit a power-law to the $\rho_{\rm UV}$ values out to z$=2.6$, which is defined as follows:
\begin{equation}
    \rho_{\rm UV} = N~\times(1+z)^{\beta},
\end{equation}
\noindent where $\beta$ is the power-law exponent and $N$ is the normalization constant. The fit is carried out using orthogonal distance regression \citep{Boggs90}, accounting for error in $\rho_{\rm UV}$ and the redshift bin-width as the uncertainty in z. The best fit power-law is shown in Figure~\ref{fig:lumden} with $N=25.34\pm0.06$~erg/s/Hz/Mpc$^{3}$ and $\beta=2.46\pm0.2$. This is consistent within errors with $\beta=2.58\pm0.15$ determined by \citet{oesch10}, but steeper than $\beta=1.74$ determined by \citet{Alavi16}, as evident in Figure~\ref{fig:lumden}. 

\citet[][hereafter MD14]{MD14} defined a modified double-power law function to describe star-formation rate density over the entire range of redshifts. This function was modified by \citet{Alavi16} to instead describe the equivalent $\rho_{\rm UV}$ values. The function is defined as follows:
\begin{equation}
    \rho_{\rm UV} = a\times\frac{(1+z)^{b}}{1+[(1+z)/d]^{c}},
\end{equation}
\noindent where a, b, c and d are associated with the normalisation, rising exponent at low redshift, peak of $\rho_{\rm UV}$, and falling exponent at high redshift respectively. The best fit \citetalias{MD14} function is shown in Figure~\ref{fig:lumden}
with $a=25.35\pm0.07$, $b=2.44\pm0.21$, $c=6.29\pm0.36$ and $d=4.44\pm0.26$. The normalisation and rising exponent are nearly identical to that determined from the power-law fit, and consequently the \citetalias{MD14} function fit overlaps with the power-law fit at z$<2$ (see Figure~\ref{fig:lumden}). \citet{Alavi16} also fitted the \citetalias{MD14} function (see Figure~\ref{fig:lumden}) with a higher normalisation and both rising and falling exponents being less steep. Our determined power-law and \citetalias{MD14} function fits at z$<3$ are now consistent with each-other, compared to the previous determination. Through the availability of additional log($\rho_{\rm UV}$) determinations at z$>3$, we better constrain the falling exponent of the \citetalias{MD14} function with increasing redshift.
 
\section{Conclusion} 
\label{sec:conc}

Including the UVLF parameters determined in this work, we find that the log($\rho_{\rm UV}$) slope as a function of z (see Figure~\ref{fig:lumden}) at z$<$3 is steeper than the previous best determination by \citet{Alavi16}. This implies that star-formation decline from cosmic dawn to present day was more rapid than the previous estimate by \citet{Alavi16}. We find consistent power-law and \citetalias{MD14} functional fits of the log($\rho_{\rm UV}$) trend with redshift at z$<$3. At z$>$3, by including the literature UVLF parameters obtained out to z$\sim$10, we find the falling log($\rho_{\rm UV}$) slope is significantly steeper than previous determinations. This implies that star-formation increase till cosmic dawn was also more rapid than the previous estimates.

Cosmic star-formation thus peaked at z$\sim$3 (see peak of \citetalias{MD14} functional fit in Figure~\ref{fig:lumden}) where-after star-formation has decreased till its present quiescent phase \citep[see also][and references therein]{Forster20}. To better probe the presence of any short-lived global fluctuations of star formation superimposed on its slow decline since cosmic noon, the UVLF needs to be determined with similarly fine high redshift bins, as achieved in this work, in other observed patches of the universe and at even higher redshifts.


\section*{Acknowledgements}

We thank the anonymous referee for their comments. This work is primarily based on observations taken by AstroSat/UVIT. The UVIT project is a result of collaboration between IIA, Bengaluru, IUCAA, Pune, TIFR, Mumbai, several centres of ISRO, and CSA.  Several groups from ISAC (ISRO), Bengaluru, and IISU (ISRO), and Trivandrum have contributed to the design, fabrication, and testing of the payload. The Mission Group (ISAC) and ISTRAC (ISAC) continue to provide support by making observations with, and reception and initial processing of the data. This research made use of Astropy-- a community-developed core Python package for Astronomy \citep{Rob13}, SciPy \citep{scipy}, NumPy \citep{numpy} and Matplotlib \citep{matplotlib}. This research also made use of NASA’s Astrophysics Data System (ADS\footnote{\url{https://ui.adsabs.harvard.edu}})

\section*{Data Availability}

The AUDFs source catalogue used in this work has already been published by \citet{Saha24}. The HDUV survey catalogue has been published by \citet{Oesch18}. The CANDELS photometric redshift catalogue has been published by \citet{Kodra23}.



\bibliographystyle{mnras}
\bibliography{uvit} 


\bsp	
\label{lastpage}
\end{document}